# RF SYSTEM UPGRADES TO THE ADVANCED PHOTON SOURCE LINEAR ACCELERATOR IN SUPPORT OF THE FEL OPERATION[*]

T.L. Smith[†], M.H. Cho[+], A.E. Grelick, G. Pile, A. Nassiri, N. Arnold

Argonne National Laboratory, Argonne, IL60439 USA

*Abstract*

The S-band linear accelerator, which was built to be the source of particles and the front end of the Advanced Photon Source [1] injector, is now also being used to support a low-energy undulator test line (LEUTL) and to drive a free-electron laser (FEL). The more severe rf stability requirements of the FEL have resulted in an effort to identify sources of phase and amplitude instability and implement corresponding upgrades to the rf generation chain and the measurement system. Test data and improvements implemented and planned are described.

## 1 INTRODUCTION

The rf power for the APS linear accelerator [2] is provided by five klystrons (L1 through L5), each of which feeds one linac sector; L1 feeds rf power to a thermionic rf gun via the exhaust of one accelerating structure (AS). The L2, L4, and L5 sectors are conventional sectors, each using a SLED cavity assembly [3] to feed four ASs. L3 supplies rf power to the photocathode gun located at the beginning of the linac. For normal storage ring injection operation, L1, L2, L4, and L5 are operated; for self-amplified spontaneous emission (SASE)-FEL operation, all units are operated. A pulsed solid-state amplifier is used to drive the klystrons. The pulsed amplifier is preceded by a drive line that feeds all sectors, phase shifters, a PIN diode switch used for VSWR protection, and a preamplifier. The rf for the entire linac is derived from an ovenized, synthesized source and 10-Watt GaAs FET amplifier, which feeds both the drive line and a reference line. The VXI-based measuring system for each sector is housed in a separate cabinet, and each system uses the reference line to derive phase measurements. The overall APS linac rf system is schematically shown in Fig. 1.

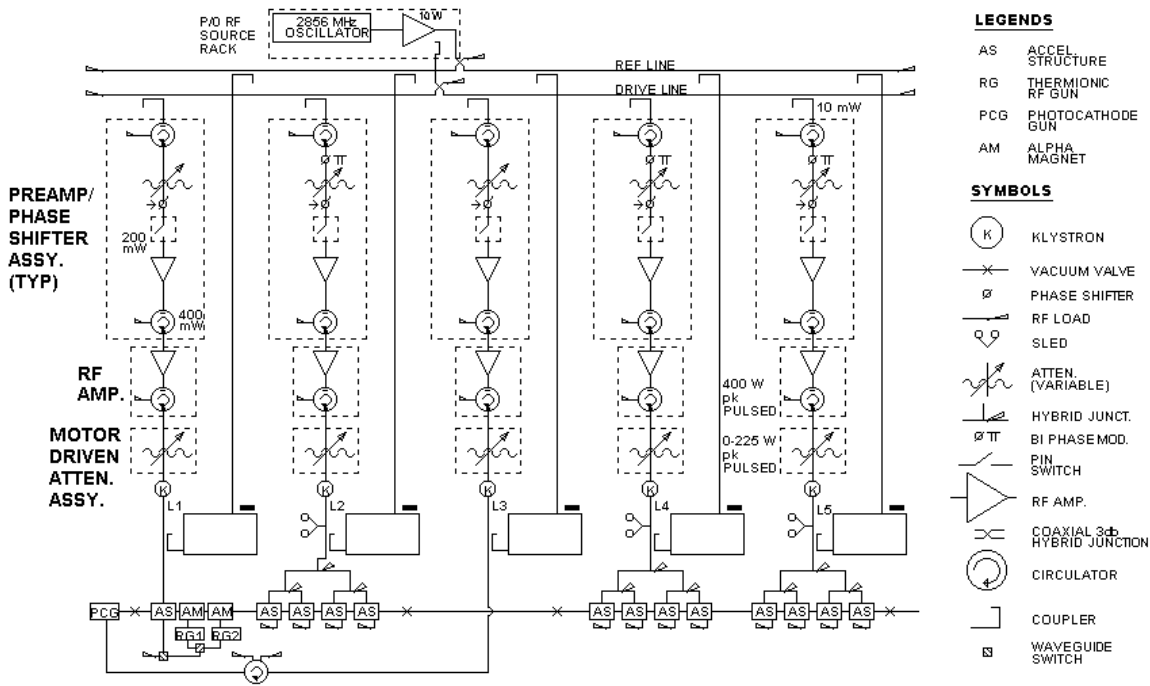

Figure 1: APS linac rf schematic diagram.

[*] Work supported by the U.S. Department of Energy, Office of Basic Sciences, under Contract No. W-31-109-ENG-38.
[†] Email; tls@aps.anl.gov
[+] Permanent address; Physics Department, POSTECH, Pohang, S. Korea, Email: mhcho@postech.ac.kr

## 2 PHASE STABILITY

### 2.1 Phase Stability Test Unit

A new phase measurement circuit (Fig. 2) was built that allowed two 2856-MHz signals to be input to a double-balanced mixer with the resultant mixer output (IF) showing a phase relationship between the two input signals. A network analyzer was used to measure the phase at 2856 MHz of a variable phase shifter for calibration over the variable phase shifter vernier scale.

For calibration of the phase stability test unit, a 2856-MHz signal generator output was split and connected to the two inputs. Both variable attenuators were adjusted so the mixer LO and RF inputs = +10 dBm. The variable phase shifter was adjusted around the zero crossing to get a calibrated reading of 1 mV = 0.219°. To make accurate calibrated linac phase measurements, LO and RF ports were always set to +10 dBm and measurements were always taken at the mixer IF output zero crossing.

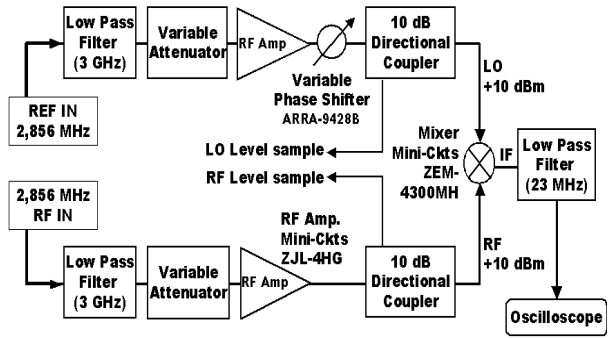

Figure 2: Block diagram of phase stability test unit.

### 2.2 Modulator Upgrades

The modulator is a classic line design with a pulse forming network (PFN) and thyratron switch. The DC high-voltage power supply section charges the PFN to a voltage of up to 35 kV before each pulse. When the modulator is triggered, the PFN is discharged by the thyratron (EEV1836A) through the primary of a pulse transformer, resulting in a high-power pulse on the secondary (approx. –300 kV for 5 µs) being applied to the klystron cathode. Precise regulation of the PFN charging voltage is desired, since the klystron phase and amplitude are sensitive to this voltage. Original modulator power supplies had 3-phase variable transformers for the incoming AC voltage control, which are ineffective for high speed regulation. The electromechanical regulator plus the series-connected command-charging switch combination also did not satisfy our goal of better than 99% availability.

All modulators have been upgraded and now use a constant current mode, solid state, high frequency inverter type high-voltage power supply (EMI-303L). This power supply provides better than ±0.3% charging voltage regulation performance [4], improving klystron output phase stability. The phase stability measured across the klystrons has improved by a factor of 2 to 2.5.

### 2.3 Adjustment of Phase Compensation Circuit

The low-level pre-amp phase shifter assembly operates on a 2856-MHz CW signal with the add-on function of phase compensation during the klystron drive pulse. The compensation is designed to cancel out rather slow phase settling of the 400-W amplifier output pulse that is due to the thermal time constant of the pulsed output transistors within the amplifier. The leading edge of the rf gate triggers an adjustable R-C circuit, providing an exponential waveform that is applied to a dedicated compensation phase shifter [5]. The practice has been to adjust the compensation for the flattest drive phase after settling of a leading edge transient. It has been determined that adjusting for best phase at the klystron output can improve system performance. Figures 3 and 4 show the L3 klystron phase before and after the compensation adjustment.

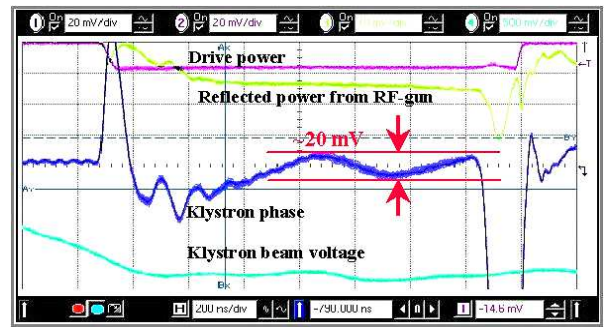

Figure 3: L3 Klystron phase – before adjustment of phase compensator.

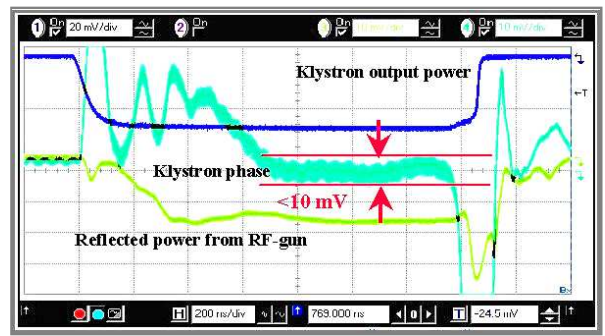

Figure 4: L3 Klystron phase – after adjustment of phase compensator.

### 2.4 Modulator Trigger Adjustments

A typical "phase pushing" specification from the klystron vendor (Thomson) could be approx. 600 V = 1° of phase shift. This varies with the cathode voltage used and the unique characteristics of each klystron. The modulator trigger could be adjusted so that the flattest section of the cathode voltage is present during the time when the beam is present. This would produce a more stable phase. Figures 5 and 6 show data before and after modulator adjustments.

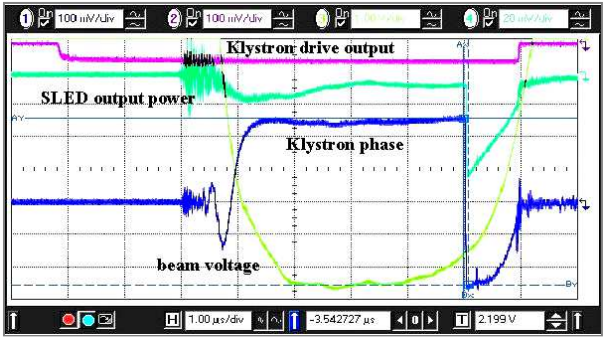

Figure 5: L2 before modulator trigger adjustments.

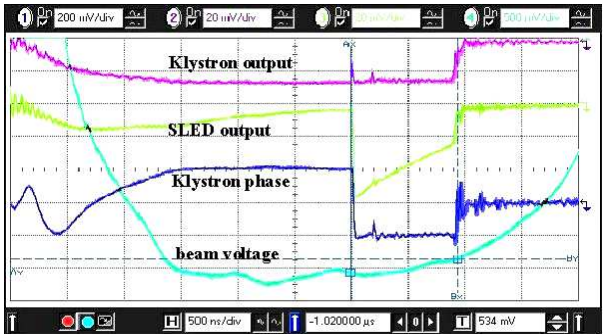

Figure 6: L2 after modulator trigger adjustments.

## 2.5 Trigger Jitter

Each modulator-klystron subsystem receives three modulator triggers, an rf drive gate, a SLED PSK trigger, and a klystron pulse trigger. Each trigger has unique timing requirements that are functions of trigger distribution delays, internal delays, and the beam transmission time. A number of commercial precision delay generators (SRS:DG535) and VXI-based delayed trigger generators are the main hardware units for the trigger system.

An approximately 10-ns trigger jitter has been observed, for example, between the rf drive and the SLED PSK trigger during normal injection operation to the booster ring. However, the requirement of the phase stability for the current on-going SASE-FEL experiment (LEUTL [6]) is less than $2°$, which is equivalent to 2-ps in time. An experimental study is planned to explore the possibility of satisfying the above requirement, without replacing the existing trigger delay system, by improving the flat-top of the klystron output power characteristics.

## 3 ON-LINE PHASE MEASUREMENT SYSTEM

Phase is measured online using reference and sample signals that are synchronously downconverted to 20 MHz. A vector detector operating at 20-MHz produces I and Q vector signals from which phase is calculated using software [7].

Each existing I/Q demodulator is implemented with analog circuits utilizing two double-balanced mixers in an insulated, ovenized enclosure. Because 10 MHz bandwidth is required to follow SLED waveforms accurately, there is a 20 MHz component that cannot be completely filtered out and constitutes noise in the measurements because it is asynchronous to the pulse timing and hence the sample in each pulse.

The planned upgrade will use a 12-bit flash A-to-D converter sampling synchronously to the 20-MHz IF at 80-MSa/s, to allow retrieval of data unaffected by carrier noise. Together with a reduction of LO noise, achieved by upgrading to a newer IF reference oscillator design, it is expected that the repeatability of the relative phase value in consecutive readings will be within a fraction of one degree. This should allow the use of feedback to reduce phase perturbations.

## 4 FUTURE IMPROVEMENTS

Linear accelerator operation for a SASE FEL, the $4^{th}$ generation light source, requires state-of-the-art performance from the all linac subsystems, especially rf power (within ±0.1% shot-to-shot) and phase (less than $1°$ rms jitter shot-to-shot and $±5°$ rms long-term drift). To satisfy such tight criteria, both low-level and high-power rf systems are under evaluation together with development of a precision on-line phase detector system. Along with the upgrade of relevant hardware components and subsystems, a beam-based optimization study of the operation parameters, such as trigger and phase compensation adjustment, is also planned.